\documentclass[fleqn,twoside]{article}
\usepackage{espcrc2}

\usepackage{graphicx}
\usepackage[figuresright]{rotating}

\newcommand{\AmS}{{\protect\the\textfont2
  A\kern-.1667em\lower.5ex\hbox{M}\kern-.125emS}}

\title{Spin order manipulations in nanostructures of II-VI ferromagnetic semiconductors}

\author{Tomasz Dietl\address{Institute of Physics, Polish Academy of Sciences\\ al. Lotnik\'ow 32/46, PL-02668 Warszawa, Poland}%
        \thanks{Address in 2003: Institute of Experimental and Applied Physics, Regensburg University, Germany; 
supported by Alexander von Humboldt Foundation; e-mail: dietl@ifpan.edu.pl.}}
       
\begin{document}

\begin{abstract}
An overview of recent studies on ferromagnetism in Cr- and Mn-based II-VI diluted magnetic semiconductors is presented emphasizing differences in underlying exchange mechanisms. Examples of manipulations with spin ordering by carrier density, dimensionality, light, and electric field are given.
\vspace{1pc}
\end{abstract}

\maketitle

\section{INTRODUCTION}

Present spintronic research \cite{Ohno02,Wolf01,Diet01c} involves virtually all materials families but ferromagnetic semiconductors are particularly attractive as they combine resources of both semiconductors and ferromagnets. We describe here recent studies of ferromagnetic Cr- and Mn-based II-VI diluted magnetic semiconductors (DMS) excluding, however, oxides that are subject of another presentation \cite{Jeon03}. Properties of III-V DMS are summarized elsewhere \cite{Diet02,Mats02,Ohno03}; detail reviews of nitride DMS have also been recently completed \cite{Pear03,Diet03}. 

\begin{figure}
\includegraphics[width=0.5\textwidth]{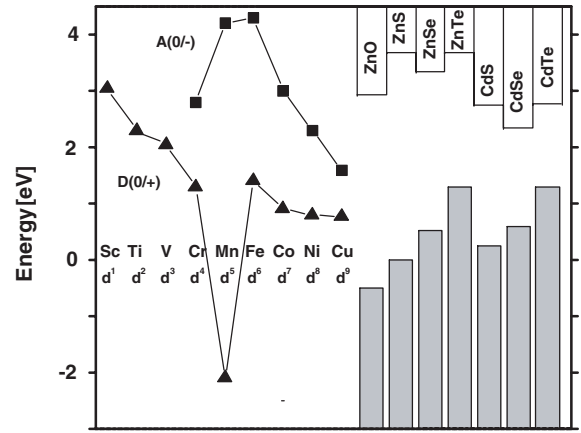}
\caption{Approximate positions of transition metals levels
relative to the conduction and valence band edges of II-VI compound semiconductors. By triangles the
d$^{N}$/d$^{N-1}$ donor and by squares the d$^N$/d$^{N+1}$ acceptor
states are denoted (adapted from Ref.~\cite{Zung86,Lang88,Blin02}).}
\label{levels_II_VI}
\end{figure}

A good starting point for the description of DMS is the Vonsovskii model, which assumes that the electronic structure consists of extended sp band states and highly localized d-shells of magnetic ions \cite{Diet94}. As depicted in Fig.~\ref{levels_II_VI}, the positions of the lower and upper Hubbard levels in respect to band edges are universal if the relative positions of the band edges are shifted according to band offsets known from heterostructure studies \cite{Zung86,Lang88}. This diagram makes it possible to asses the ion charge state and its variations with co-doping by shallow impurities. For a given charge state, the ion spin is determined by Hund's rule. 

\begin{figure*}[t]
\includegraphics[width=0.9\textwidth]{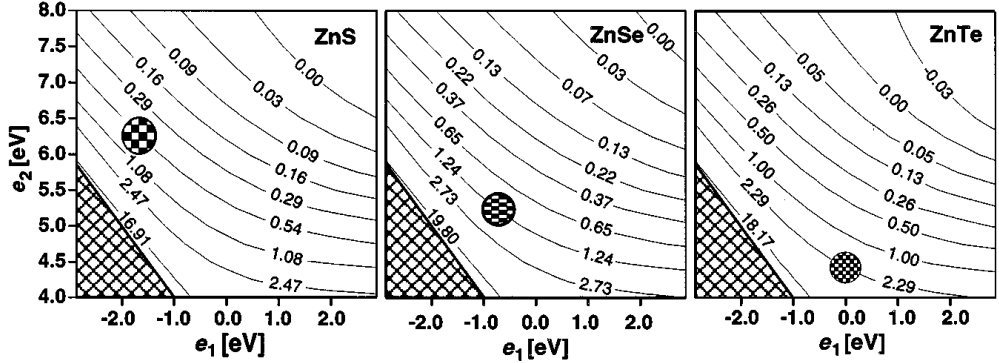}
\caption{Contour graphs showing the dependence of a superexchange ferromagnetic energy $J$ (in K) on the d-level energies $e_1$ and $e_2$ for the nearest-neighbor Cr pair with one of possible directions of Jahn-Teller distortions in ZnS, ZnSe, and ZnTe assuming Harrison's $V_{pd\sigma} = 1$~eV. Here, $e_1$ denotes the position of the valence band edge in respect to the D(0/+) level of Fig.~1 and $e_2$ is the A(0/-) energy for the total spin $S=3/2$ in respect to the valence band edge. The shaded circles define approximately the areas of the $e_1$ and $e_2$ values compatible with the positions of the Cr levels of Fig.~1 assuming the difference between the excited $S = 3/2$ and the ground state $S = 5/2$ to be $\Delta = 3$~eV. The cross-hatched triangles indicate the nonphysical region of $e_1$ and $e_2$ values, where the D(/+) donor lies below A(0/-) acceptor, {\it i.e.},  $e_2-\Delta +e_1<0$ (after Ref.~\cite{Blin96a}).}
\label{Blinowski_Cr}
\end{figure*}

\section{FERROMAGNETISM IN Cr-BASED DMS}

Owing to short-range superexchange interactions, the spin-spin coupling is merely antiferromagnetic in II-VI DMS. However, a net ferromagnetic superexchange was predicted for Cr- \cite{Blin96a} and V-based \cite{Blin96b} II-VI DMS, a conclusion independent of assumptions concerning the direction of Jahn-Teller distortion and tight-binding parameter values, as shown in Fig.~\ref{Blinowski_Cr}. A ferromagnetic ground state in these systems is expected also from more recent {\em ab initio} computations \cite{Sato02} which suggest, however, that double exchange rather than superexchange is involved. 

\begin{figure}[t]
\includegraphics[width=0.45\textwidth]{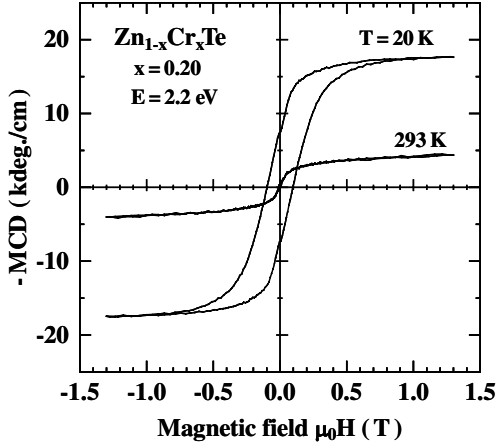}
\caption{Magnetic field dependence of magnetic circular dichroism in Zn$_{1-x}$Cr$_x$Te ($x=0.20$) film at photon energy $E=2.2$~eV and temperature of 20~K and 293~K indicating the presence of a ferromagnetic ordering persisting up to room temperature (after Ref.~\cite{Sait03}).}
\label{Ando_Cr}
\end{figure}

In the initially studied samples \cite{Mac96,Wojt97}, the Cr concentration was too low to tell the character of spin-spin interactions. However, a more recent work on (Zn,Cr)Te \cite{Sait02a,Sait02b,Sait03},  has lead to the observation ferromagnetism by both direct magnetization measurements and magnetic circular dichroism (MCD).  According to MCD results presented in Fig.~\ref{Ando_Cr}, ferromagnetism persists up to room temperature in a sample containing 20\% of Cr. Since no electrical conductance is detected, the ferromagnetic double exchange \cite{Sato02} does not appear to operate. At the same time, for the presently accepted values of the parameters, the observed magnitude of $T_{\mbox{\tiny{C}}}$ is too high to be explained within the superexchange scenario but by adjusting parameters within the physically acceptable range, such a scenario becomes plausible. Nonetheless, independently of the microscopic nature of ferromagnetic ordering, the large magnitude of MCD suggests possible applications of this system in photonic devices, such as Faraday optical insulators.

\begin{figure}[t]
\includegraphics[width=0.4\textwidth]{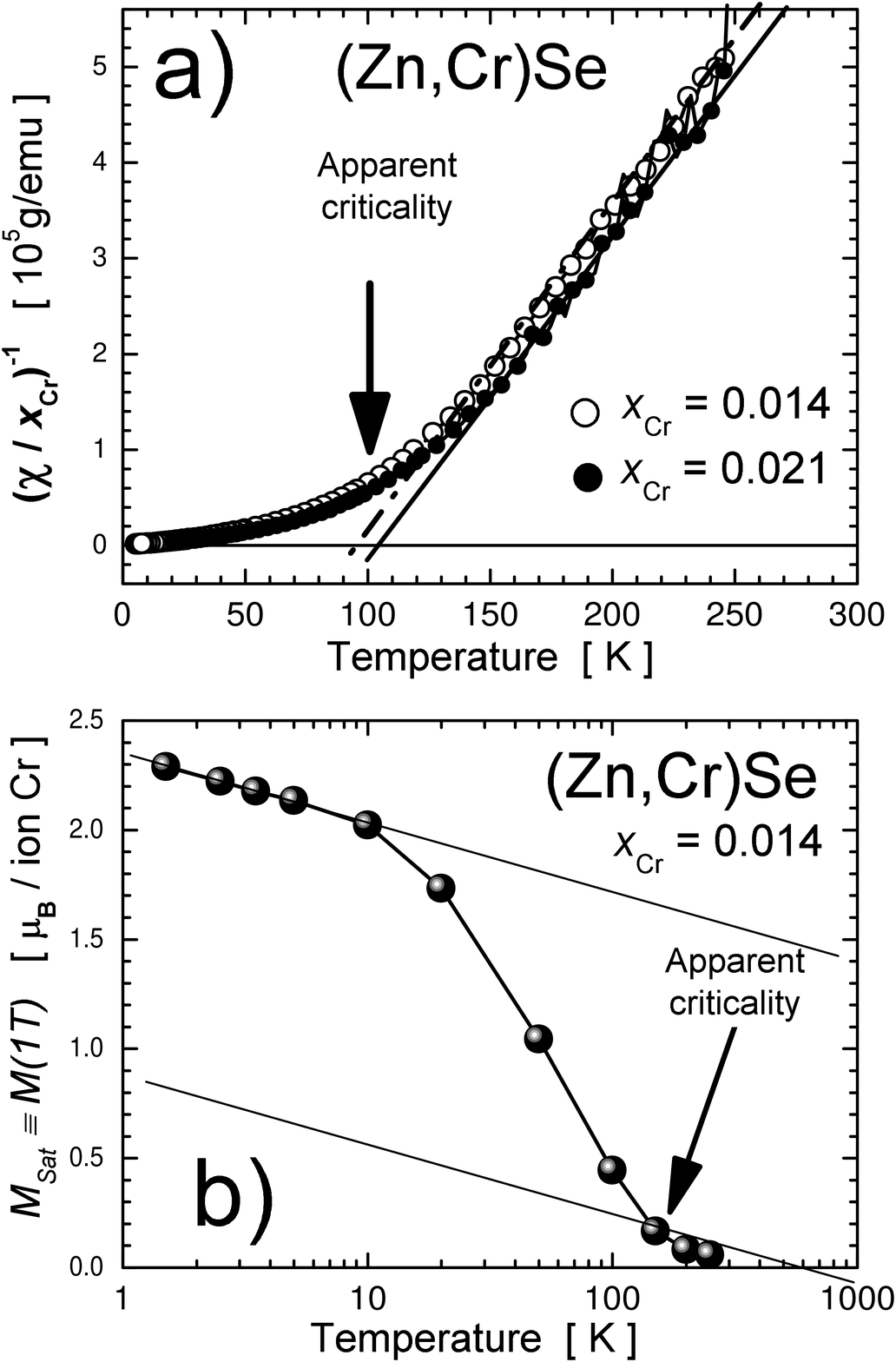}
\caption{Temperature dependence of inverse magnetic susceptibility (a) and magnetization at 1~T (b) for Zn$_{1-x}$Cr$_x$Se with $x=0.014$ and $x =0.021$. The appearing cross-over to a ferromagnetic phase below 100~K is assigned to precipitates of ZnCr$_2$Se$_4$ (after Ref.~\cite{Karc03}).}
\label{Sawicki_Cr}
\end{figure}

Indications of ferromagnetism below 100~K have also been found in (Zn,Cr)Se, as shown in Fig.~\ref{Sawicki_Cr}.  The enhanced magnetic response has been assigned to precipitates as the apparent $T_{\mbox{\tiny{C}}}$ does not scale with the Cr concentration, and is close to $T_{\mbox{\tiny{C}}}$ of the spinel semiconductor ZnCr$_2$Se$_4$.

\section{FERROMAGNETISM IN Mn-BASED DILUTED MAGNETIC SEMICONDUCTORS}

\begin{figure}[t]
\includegraphics[width=0.4\textwidth]{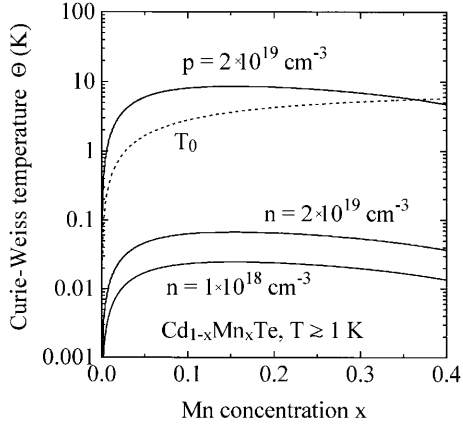}
\caption{Calculated mean-field value of the carrier-induced ferromagnetic Curie-Weiss temperature $\Theta$ 
for p- and n-type Cd$_{1-x}$Mn$_x$Te, compared to antiferromagnetic
temperature $T_o(x)$. Ferromagnetic phase transition is expected
if $\Theta > T_o$. Material parameters as determined at 1.7~K were
adopted for the calculation (after Ref.~\cite{Diet97}).}
\label{Dietl_97}
\end{figure}

According to extensive studies of Mn-based II-VI DMS, the intrinsic antiferromagnetic interaction gives rise to a spin-glass freezing at low temperatures \cite{Diet94}. It was predicted that the antiferromagnetic coupling can be overcompensated by ferromagnetic interactions mediated by the valence band holes in various dimensionality systems \cite{Diet97}, as depicted for the 3D case in Fig.~\ref{Dietl_97}. This expectation has been confirmed by experimental studies of p-type modulation-doped (Cd,Mn)Te quantum wells \cite{Haur97,Bouk02} as well as of p-type (Zn,Mn)Te:N \cite{Ferr01,Andr01,Sawi02}, (Zn,Mn)Te:P \cite{Andr01,Sawi02}, and (Be,Mn)Te:N \cite{Hans01}.

\begin{figure}[t]
\includegraphics[width=0.55\textwidth]{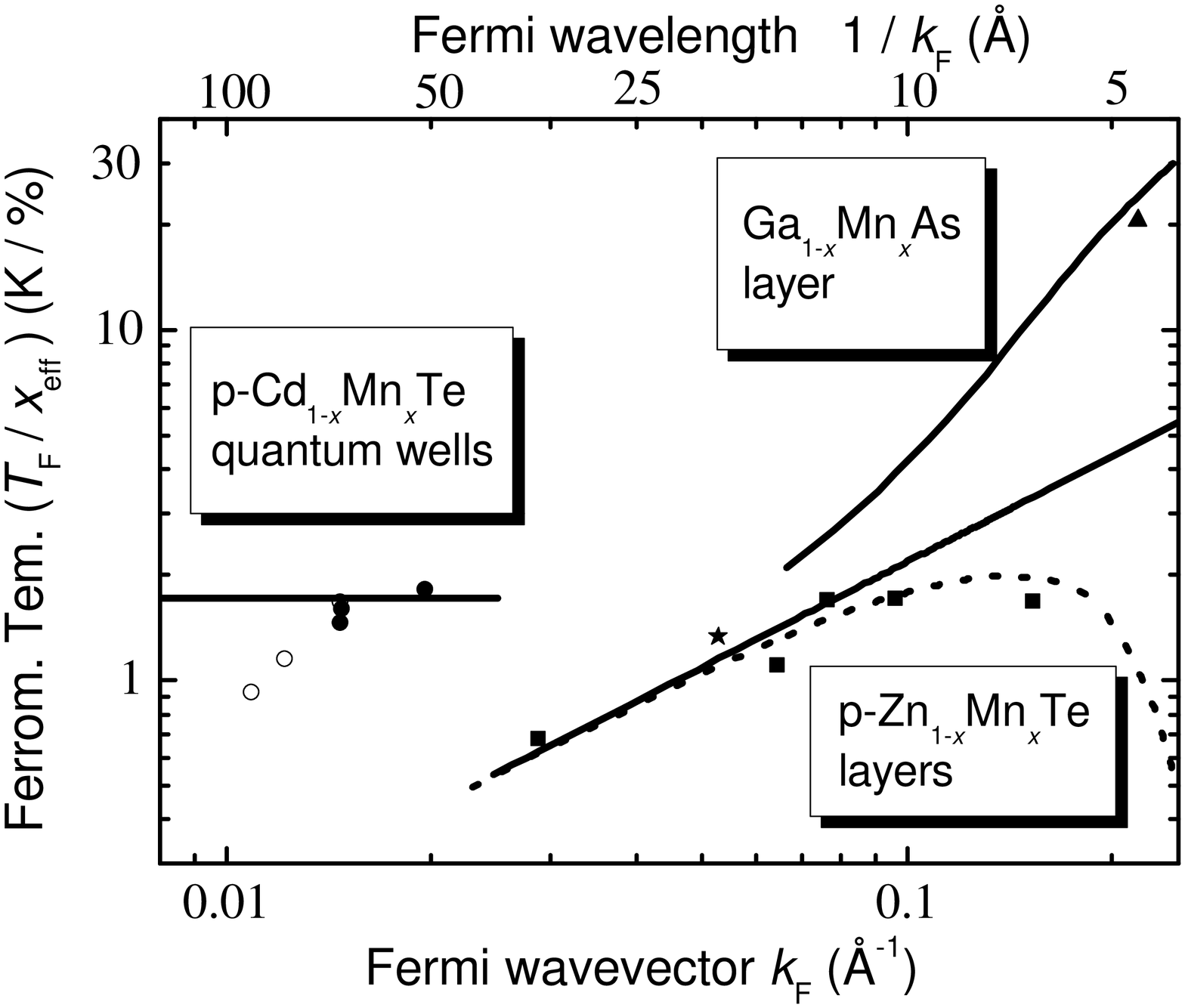}
\caption{Experimental (symbols) and calculated (lines) normalized
ferromagnetic temperature, $T_{\mbox{\small F}}/10^2x_{\mbox{\small
eff}}$, versus the wave vector at the Fermi level for
Ga$_{1-x}$Mn$_x$As (triangle) \cite{Mats98,Omiy00},
Zn$_{1-x}$Mn$_x$Te:N (squares) \cite{Ferr01,Andr01},
Zn$_{1-x}$Mn$_x$Te:P (star) \cite{Andr01,Sawi02}, and quantum well of
p-Cd$_{1-x}$Mn$_x$Te (circles) \cite{Koss00,Cibe02}.  Solid lines:
Zener and $6\times6$ Luttinger model for the 3D  \cite{Diet01})
and 2D case \cite{Diet97}; dotted line: the RKKY and $6\times6$
Luttinger model for $x_{\mbox{\small eff}}=0.015$, taking into
account the effect of the antiferromagnetic interactions on
statistical distribution of unpaired Mn spins \cite{Ferr01}.}
\label{Cibert}
\end{figure}

On the level of the mean-field and continuous medium approximations, the RKKY approach is equivalent to the Zener model \cite{Diet97}. In terms of the latter, the equilibrium magnetization, and thus $T_{\mbox{\tiny{C}}}$ is determined by minimizing the Ginzburg-Landau free energy functional $F[M(r)]$, where $M(r)$ is the local magnetization of the localized spins \cite{Diet00,Diet01}. This is a rather versatile approach, to which carrier correlation, confinement, $k\cdot p$, and spin-orbit couplings as well as weak disorder and antiferromagnetic interactions can be introduced in a controlled way. Importantly, by evaluating $F[M_q]$ for the carriers, the magnetic stiffness can be determined \cite{Koen01}. As shown in Fig.~\ref{Cibert}, theoretical calculations \cite{Ferr01,Diet01}, carried out with no adjustable parameters, explain satisfactory the magnitude of $T_{\mbox{\tiny{C}}}$ in both p-type (Zn,Mn)Te \cite{Ferr01} and (Ga,Mn)As \cite{Mats98,Omiy00}. A question arises how the character of the spin-spin interactions changes if the hole concentration becomes so small that the material is on the insulator side of the metal-to insulator transition (MIT). According to the scaling theory of the Anderson-Mott MIT at distances smaller than the localization radius (which diverges right at the MIT), the states retain a metallic character, so that the Zener/RKKY appears valid \cite{Diet97,Diet00}. Recent studies of the interaction energy of the nearest neighbor Mn pairs by inelastic neutron scattering in (Zn,Mn)Te:P \cite{Kepa03} corroborate this conjecture.  Owing to relatively small magnitudes of the s-d exchange coupling and density of states, the carrier-induced ferromagnetism is expected \cite{Diet97}, and observed only under rather restricted conditions in n-type Mn-based DMS \cite{Andr01,Jaro02}.

According to the above model, $T_{\mbox{\tiny{C}}}$ is proportional to the density of states for spin excitations, which is independent of energy in the 2D systems.  Hence, $T_{\mbox{\tiny{C}}}$ is expected to do not vary with the carrier density, and to be enhanced over the 3D values at low carrier densities. Experimental results for modulation doped p-type (Cd,Mn)Te quantum wells \cite{Haur97,Bouk02}, presented in Fig.~\ref{Cibert}, fulfill these expectations, though a careful analysis indicates that disorder-induced band tailing lowers $T_{\mbox{\tiny{C}}}$ when the Fermi energy approaches the band edge \cite{Bouk02,Koss00}. In 1D systems, in turn, a formation of spin density waves with $q = 2k_F$ is suggested, a prediction awaiting for an experimental confirmation. A good agreement between experimental and theoretical values of $T_{\mbox{\tiny{C}}}$ in (Ga,Mn)As, p-(Cd,Mn)Te, and p-(Zn,Mn)Te has triggered the extension of the calculations to other Mn-based III-V, II-VI, and IV systems \cite{Diet00,Diet01}.

\section{MAGNETIZATION MANIPULATION}

\begin{figure}[t]
\includegraphics[width=0.5\textwidth]{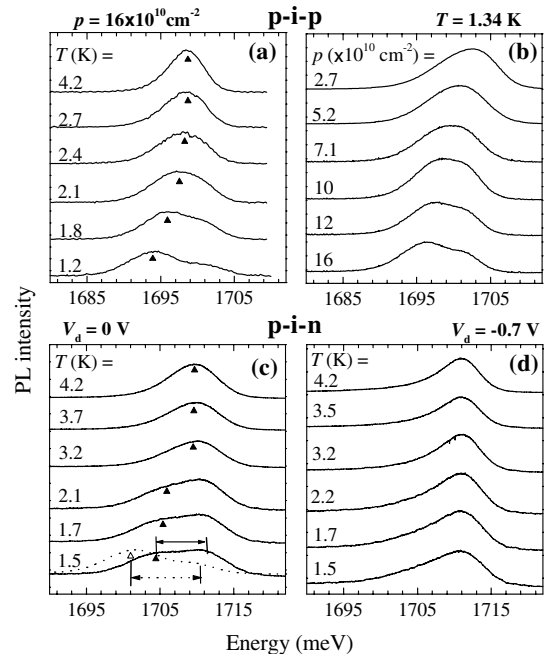}
\caption{Effect of temperature (a,c,d), illumination (b) and bias voltage $V_d$ (c,d) on photoluminescence line in quantum well of (Cd,Mn)Te placed in a center of p-i-n diode (c,d) and p-i-p structure (a,b). Line splitting and shift witness the appearance of a ferromagnetic ordering that can be altered by light (b) and voltage (c,d), which change the hole concentration $p$ in the quantum well (after Ref.~\cite{Bouk02}).}
\label{Boukari}
\end{figure}

Since magnetic properties of the Mn-based compounds are controlled by the band carriers, the powerful methods developed to change carrier concentration by electric field and light in semiconductor quantum structures can be employed to alter the magnetic ordering. As far as II-VI DMS are concerned, such tuning capabilities were put into the evidence in (Cd,Mn)Te quantum wells  \cite{Haur97,Bouk02}, as shown in Fig.~\ref{Boukari}.  Importantly, the magnetization switching is isothermal and reversible. Though not investigated in detail, it is expected that underlying processes are rather fast. Since the background hole concentration is small in Mn-based II-VI quantum wells, the relative change of the Curie temperature is typically larger than in III-V compounds.

\section{OUTLOOK}

The case of ferromagnetism  research in Cr-based and Mn-based II-VI DMS indicates the important role of sensible theoretical modelling, provided that it is accompanied by parallel developments of growth modes which can overcome solubility and self-compensation limits as well as prevent the formation of precipitates.

\section*{ACKNOWLEDGMENTS}
The author would like to thank his co-workers, particularly P. Kacman, P. Kossacki, and M. Sawicki in Warsaw, F. Matsukura and H. Ohno in Sendai, J. Cibert in Grenoble, and A.H. MacDonald in Austin for many years of fruitful collaboration. Author's research was supported by State Committee for Scientific Research, by AMORE (GRD1-1999-10502) and FENIKS (G5RD-CT-2001- 00535) EC projects, and by Ohno Semiconductor Spintronics ERATO Project of JST.

\end{document}